\documentclass[floatfix,
reprint,
superscriptaddress,
 amsmath,amssymb,
 aps,
pra,
]{revtex4-2}

\usepackage{gensymb}
\usepackage{graphicx}
\usepackage{dcolumn}
\usepackage{bm}
\usepackage[utf8]{inputenc}

\usepackage{graphicx}
\usepackage{physunits}
\usepackage{xcolor}
\usepackage{amsmath}
\usepackage{multirow}
\usepackage{ulem}
\begin{document}

\preprint{APS/123-QED}

\author{Melissa Yactayo}
\email{melissa.yactayo@unmsm.edu.pe}
\affiliation{Université de Lorraine, CNRS UMR 7198, Institut Jean Lamour, Nancy, France}
\affiliation{Universidad Nacional Mayor de San Marcos, P.O.-Box 14–0149, Lima 14, Peru}
\author{Michel Hehn}
\affiliation{Université de Lorraine, CNRS UMR 7198, Institut Jean Lamour, Nancy, France}
\author{J.-C. Rojas-S\'anchez}
\email{juan-carlos.rojas-sanchez@univ-lorraine.fr}
\affiliation{Université de Lorraine, CNRS UMR 7198, Institut Jean Lamour, Nancy, France}
\author{Sébastien Petit-Watelot}
\email{sebastien.petit@univ-lorraine.fr}
\affiliation{Université de Lorraine, CNRS UMR 7198, Institut Jean Lamour, Nancy, France}

\title{RF field characterization and rectification effects in spin pumping and spin-torque FMR for spin-orbitronics}

\date{\today}

\begin{abstract}

Quantifying spin-orbital-to-charge conversion efficiency is crucial for spin-orbitronics. Two widely used methods for determining these efficiencies are based on ferromagnetic resonance (FMR), spin pumping FMR for the inverse effect, and spin-torque FMR for the direct effect. A key parameter to achieve accurate quantification, especially for spin-pumping FMR, is the RF field strength, $h_{\mathrm{RF}}$. We present a comprehensive theoretical model and experimental protocol that allow a correct quantification of $h_{\mathrm{RF}}$. It was validated by extensive experimental results and it was rigorously tested across various antennas geometries and ferromagnetic systems. We demonstrate that odd-symmetric Lorentzian voltages—which perfectly mimic spin-pumping or spin-torque FMR signals—can arise purely from rectification effects (due to anisotropic magnetoresistance) when $h_{\mathrm{RF}}$ orientation is parallel to the ferromagnetic surface. Through a systematic study of various 10-nm-thick ferromagnetic layers, such as Ni, NiFe, Fe, and CoFeB, we find that while Fe and CoFeB exhibit minimal rectification, Ni and NiFe generate strong rectified signals that must be corrected. We further demonstrate that these rectification effects become negligible for ferromagnetic thicknesses $\leq$ 6 nm, as validated in NiFe/Pt bilayers, providing an important guideline for the design of future heterostructures.
\end{abstract}


\maketitle

Fundamental research in spintronics and its potential applications is expanding rapidly. From the fundamental point of view, we seek to understand various phenomena associated with spin current, such as their spin polarization symmetries, generation, transmission, detection, and conversion to charge current. In orthogonal  or ``conventional" spin Hall effect (SHE) symmetry, a charge current $J_c$ produces a transverse spin current $J_s$ whose electrons have a spin polarization $\sigma$ perpendicular to $J_c$ and $J_s$, \textit{i.e.} $\mathbf J_c \propto \mathbf J_s \times \bm \sigma $. We observe this in 5d transition metals with strong spin-orbit coupling and cubic crystallographic structure, such as Pt, Ta, and W \cite{Rojas-Sanchez2019}. We also observe such symmetry in Rashba interfaces and topological insulators with orthogonal spin-momentum locking. The efficient interconversion between the spin current and the charge current allows the proposal of several technological applications such as MRAM memories, unconventional computing, or logic devices, among others \cite{Sinova2015, Manchon2019_RMP}. Therefore, quantifying the efficiency of such interconversion is crucial, as it helps separate possible spurious effects that may lead to misinterpretation of the results and misleading conclusions. 

One of the most widely used methods to study and quantify the efficiency of such interconversion is the so-called spin-pumping ferromagnetic resonance (SP-FMR) \cite{Tserkovnyak2002, Ando2011, Azevedo2011, Feng2012, Mosendz2010c, Castel2012, Rojas-Sanchez2013, rojas2014spin, Rojas-sanchezPRL2016, Lesne2016, Tao2018, Rojas-Sanchez2019, Fache2020, Arango2023Spin-to-chargeBixSe1-x, gudin2023isotropic, Ampuero2024Self-inducedFilms, Anadon2025, Sahoo2025}. Here, the resonance condition is exploited to inject a spin current from the ferromagnetic (FM) layer into the neighboring layers, such as heavy metal (HM) or bidimensional systems. This spin current is converted into a charge current in the adjacent layers if there is a finite spin-orbit coupling. The conversion may be due to the inverse spin Hall effect (ISHE) \cite{Ando2011, Azevedo2011, Feng2012, Mosendz2010c, Castel2012, Rojas-Sanchez2013, rojas2014spin, Tao2018, Rojas-Sanchez2019, Fache2020, Arango2023Spin-to-chargeBixSe1-x, gudin2023isotropic, Ampuero2024Self-inducedFilms, Anadon2025, Sahoo2025} or the inverse Edelstein effect (IEE) \cite{Rojas-sanchez2013_Ag-Bi, Rojas-sanchezPRL2016, Lesne2016, Rojas-Sanchez2019, Anadon2025}. The charge current produced is detected by measuring the voltage at the resonance condition. In the following, we will call this voltage the spin pumping voltage. The FMR condition is also exploited to study the direct effect, charge current conversion into spin current. This is also a widely used method so-called, among other names, spin-torque FMR (ST-FMR) \cite{Liu2011a, Fang2011, Kondou2012, gudin2023isotropic, Saglam2018, Guillemard2018Charge-spinDirections, Liu2019, CespedesBerrocal2021, Damas2022, gudin2023isotropic, Damas2025, Okada2019}. Here, the injected AC charge current, of the order of GHz, generates an accumulation or transverse spin current that is absorbed by the FM layer. Consequently, the injected $J_c$ generates a torque on the magnetization of the FM layer, altering its dynamics. The detection is done similarly, by measuring a rectified voltage (DC). 

In orthogonal or conventional SHE systems, the voltage from charge–spin interconversion is a symmetric Lorentzian around the resonance field $H_{\mathrm{res}}$ and is odd with respect to the external DC field, $H$, i.e., $V_{\mathrm{sym}}(H) = -V_{\mathrm{sym}}(-H)$. However, other effects can produce rectified signals with the same symmetry as spin-pumping or spin-torque FMR voltages \cite{Mecking2007,Azevedo2011,Harder2011,Rojas-Sanchez2013,Harder2016,Karimeddiny2020a,Okada2019}.
The RF magnetic field strength, $h_{\mathrm{RF}}$, depends on both the sample and the experimental setup, including CPW geometry, cavity, and substrate, making its accurate experimental determination essential.
Here, we study various antenna geometries, develop an analytical expression for $h_{\mathrm{RF}}$, and validate it experimentally in single NiFe layers and Pt/NiFe bilayers. Our focus is not on quantifying spin–orbitronic parameters—which is left for future work—but on enabling their correct determination using this protocol. We also analyze the different rectification contributions for 10 nm FM layers.

In the SP-FMR experiment, the spin current injected at the FM/HM interface towards the HM layer is determined as follows \cite{Tserkovnyak2002, Ando2011, Azevedo2011, Mosendz2010c}:
\begin{equation} \label{eq:Jsvect}
    <\mathbf{J}_S^{\Vec{n}}>=\frac{\hbar}{4\pi}\frac{1}{M_s^2}\omega g^{\uparrow\downarrow}_{eff}<\Vec{m} \times \frac{d\Vec{m}}{dt}>
\end{equation}
where $\vec{n}$ represents the normal direction of the FM/HM interface pointing toward the HM layer, $\hbar$ is the reduced Planck constant, $\omega=2\pi f$ is the microwave pulsation, $g^{\uparrow\downarrow}_{eff}$ represents the real part of the effective spin mixing conductance, $M_s$ is the FM layer magnetization saturation, $\vec{m}=\frac{\vec{M}}{M_s}$ is the normalized magnetization vector, and 
the direction of the vector $<\Vec{m} \times \frac{d\Vec{m}}{dt}>$ defines the polarization direction of the spins electrons carried by the spin current. It also reads \cite{Ando2011,rojas2014spin}:
\begin{equation}
\label{eq:Js}
J_{s}=\frac{2e}{\hbar}\frac{g^{\uparrow\downarrow}_{eff}\gamma^2 (\mu_0 h_{RF})^2 \hbar}{8\pi\alpha^2}\frac{\gamma\mu_0 M_{eff}+\sqrt{(\gamma\mu_0M_{eff})^2+4\omega^2}}{(\gamma\mu_0 M_{eff})^2+4\omega^2},
\end{equation}
where $e$ is the electron charge,  and $\gamma=g \mu_{\rm B} / \hbar$ is the gyromagnetic ratio. $\alpha$ stands for the magnetic constant damping, and $M_{eff}$ is the effective magnetic saturation. We divide the SP-FMR voltage amplitude (\textit{V}\textsubscript{SP-FMR}) by the total resistance of the slab $R$ \cite{Fache2020,rojas2014spin} to obtain the total charge current produced (\textit{I}\textsubscript{c}). Finally, the relationship between this spin current $J_s$ and the charge current produced due to ISHE is:
\begin{equation}
\label{eq:SPIc}
\begin{split}
    I\textsubscript{C}/W & =J_{s}\theta_{HM}l_{sf}^{HM}\tanh(t_{HM}/2l_{sf}^{HM}),
    \end{split}
\end{equation}

where $l_{sf}^{HM}$ is the HM spin diffusion length, $\theta_{HM}$ the HM effective spin Hall angle, and $t_{HM}$ is the HM thickness layer. For simplicity, here we consider a scenario dominated purely by the ISHE. If the charge current production is due to IEE, the product $\theta_{HM}l_{sf}^{HM}\tanh(t_{HM}/2l_{sf}^{HM})$ is replaced by $\lambda_{IEE}$ in Eq. \ref{eq:SPIc} \cite{Rojas-sanchez2013_Ag-Bi, Rojas-sanchezPRL2016, Rojas-Sanchez2019}. If there are several contributions, a linear combination of them can be made as shown for Ir/Fe/Gr/Pt in ref. \cite{Anadon2025}. However, our results shown in this study apply to all of the situations mentioned above.

This work is organized as follows: Section \ref{Section1} describes the CPW geometries and the origin of the detected rectified signals. Section \ref{Section2} presents the model used to derive Eq. \ref{eqhrf}, followed by the proposed protocol and its experimental validation across all devices in Section \ref{Section3}. Finally, in Section \ref{Section4}, we analyze rectified signals in single ferromagnetic layers using SPFMR and STFMR configurations.


\section{CPWS geometries}
\label{Section1}
\begin{widetext}

\begin{figure}[h]
\centering
\includegraphics[width=1\textwidth]{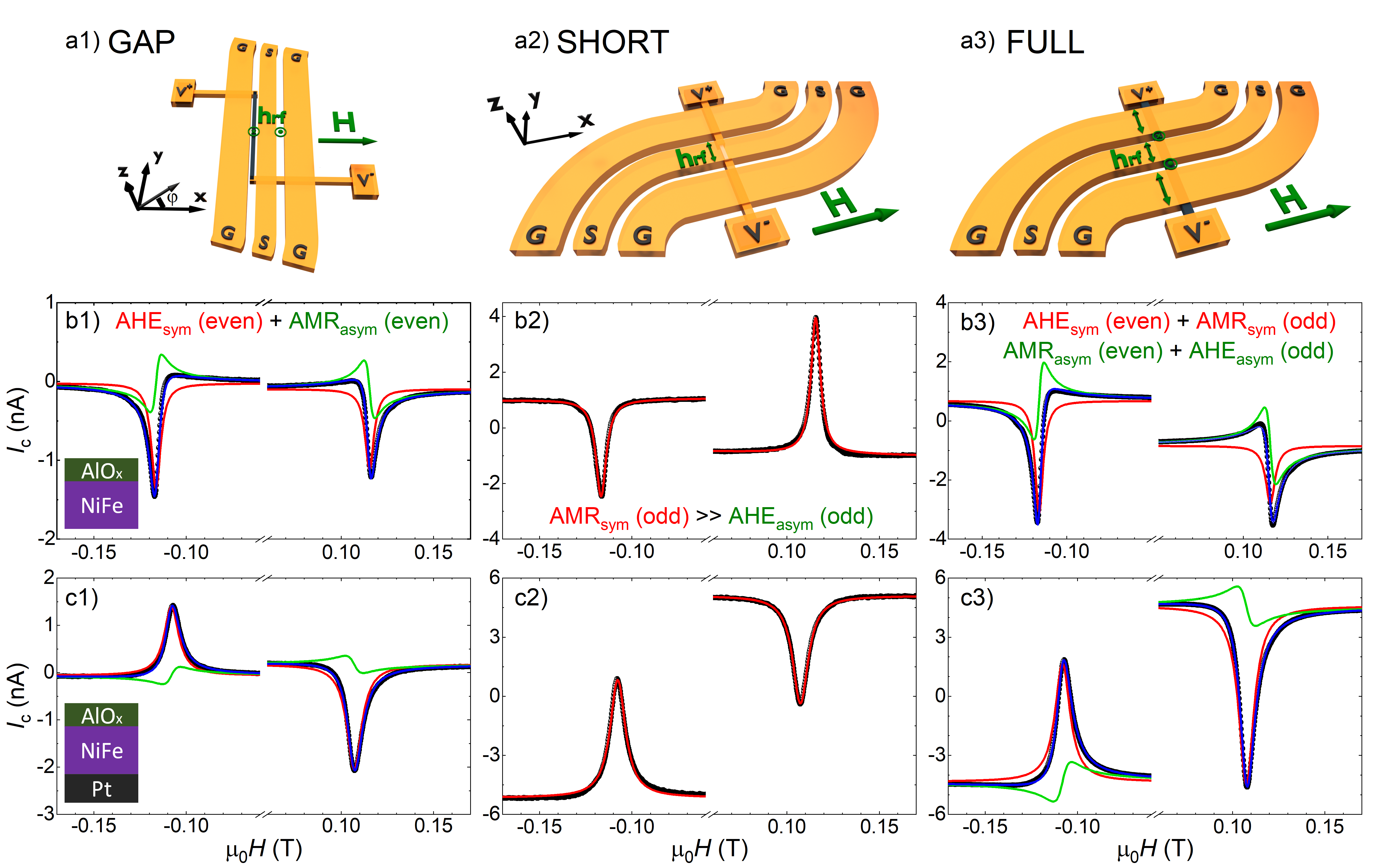}
\begin{quote}
\caption{a1–a3) 3D schematics of three GSG CPW geometries—GAP (a1), SHORT (a2), and FULL (a3)—commonly used in spin-pumping experiments. Sample width is 10 µm in all cases; lengths are 576 µm (FULL), 520 µm (GAP), and 70 µm (SHORT). b1–b3) Rectified charge current ($I_C$) versus external field for single NiFe, Si-SiO$_2$(500 nm)//NiFe(10 nm)/AlO$_x$(3 nm) at 10 GHz. GAP (b1) shows a predominantly symmetric even Lorentzian (AHE) with an antisymmetric AMR contribution; SHORT (b2) shows a symmetric odd component (AMR); FULL (b3) may include all rectified contributions. c1–c3) In //Pt(5 nm)/NiFe(10 nm)/AlO$_x$(3 nm) bilayers the signal is dominated by a symmetric odd component due to ISHE-induced charge current, as confirmed by sign reversal in SHORT (c2) compared to (b2).} \label{Fig1}
\end{quote}
\end{figure}
\end{widetext}

Figure 1 shows the three geometries of the CPW antenna used in this work to detect spin-pumping voltage. It covers two general cases: when the RF field $h_{\mathrm{RF}}$ is perpendicular to the FM/HM interface, $h_z$, and when it is parallel to both the FM/HM interface and the direction of the measured DC voltage, $h_y$. Our methodology focuses on microdevices, but the results are equally applicable to nanodevices and samples situated inside resonant cavities. The sample width is 10 µm, and the external DC magnetic field $H$ is always applied in the sample plane along the $x$ direction, which is perpendicular to the sample bar. Consequently, $H$ is always perpendicular to the direction of the measured DC voltage (measured along $y$ for all the antennas). We have named the three CPW geometries as follows:
\begin{enumerate}
    \item GAP (Fig.\ref{Fig1} a1). In this antenna, the sample is located in one of the two GAP regions of our Grounded-Signal-Grounded (GSG) microwave antenna. In this geometry, the RF field $h_{\mathrm{RF}}$ is perpendicular to the FM/HM interface of the samples, as shown in the Fig.\ref{Fig1} a1 by the green symbol.
    \item SHORT (Fig.\ref{Fig1} a2). In this microdevice, the sample is oriented perpendicularly to the GSG antenna and situated specifically under the central S-track.
    The RF field $h_{\mathrm{RF}}$ is now parallel to the long sample surface, as indicated in Fig.\ref{Fig1} a2 with the green arrow.
    \item FULL (Fig.\ref{Fig1} a3).  In this configuration, the sample geometry is similar to the SHORT design; however, it spans the entire width of the GSG antenna. Consequently, the sample connects directly to the rectangular measurement pads without requiring intermediate electrodes.   In this geometry, the $h_{\mathrm{RF}}$ field contains both parallel and perpendicular components relative to the sample surface, as also shown in Fig.\ref{Fig1} a3.
\end{enumerate}
To express the signal in current units, the measured voltage was normalized by the total resistance of the sample. For the FULL antenna, this value is consistent with the sheet resistance obtained from independent four-contact measurements. For SHORT and GAP antennas, it is normalized by the length of these samples. Namely, $R_{SHORT}=R_{FULL}\frac{70}{576}$ and $R_{GAP}=R_{FULL}\frac{520}{576}$.\\

Figures \ref{Fig1}(b1–b3) show the rectified signals measured on a single permalloy Ni$_{81}$Fe$_{19}$ (NiFe) layer, //NiFe/AlO$_x$ where $''//''$ represents the position of the substrate; and Fig. \ref{Fig1}(c1–c3) for a //Pt/NiFe/AlO$_x$ bilayer on a Si/SiO$_2$ (500 nm) substrate. Injecting GHz-range AC into the antenna induces capacitive and inductive currents along $x$, $y$, and $z$, leading to rectification via the Anomalous Hall Effect (AHE) and Anisotropic Magnetoresistance (AMR) when sweeping the external magnetic field. For a 10 nm NiFe layer, the GAP antenna shows a dominant symmetric even signal and an antisymmetric even component, which we will show are from AHE and transverse AMR ($V_{sym}^{even}$), respectively. In contrast, the SHORT antenna exhibits a primarily symmetric odd component from transverse AMR ($V_{sym}^{odd}$), which shares the same symmetry as the spin-pumping voltage. For the FULL antenna, a combination of all rectified effects is observed. In the //Pt/NiFe/AlO$_x$ bilayer, the ISHE of the Pt layer dominates across all geometries; notably, in the SHORT antenna, the symmetric signal undergoes a sign reversal compare to the NiFe(10 nm) single layer. These findings are consistent with previous reports regarding rectified voltages in various antenna configurations \cite{Mecking2007,Costache2008,Harder2011,Harder2016}.

The zero-field jump, or offset difference between positive ($H>0$) and negative fields ($H<0$), reveals the presence of thermal effects. Its prominence in SHORT and FULL antennas, contrasted with the small contribution in the GAP geometry, confirms a vertical temperature gradient due to the stacked antenna-sample architecture. We also observe a polarity reversal in this jump when comparing the bilayer to the single layer. In the case of the single layer, the jump is due to the Anomalous Nernst effect (ANE). In contrast, for the bilayer, this jump is dominated by the longitudinal spin Seebeck effect (LSSE) and the ISHE in Pt. \cite{Uchida2016,Rezende2014,Anadon2022,Palin2023}.

\section{Model}
\label{Section2}
This section gives the theoretical framework to analyze rectified voltages and to compute experimentally $h_{\mathrm{RF}}$. Further details are given in the corresponding appendix. 
In order to be as general as possible let consider the effect of basics magneto-resistance effects, that can be cast in one general vectorial relation:
\begin{equation}
   \Vec{E}=\rho_0 \Vec{J}_c + \rho_{AMR} \left(\Vec{m}.\Vec{J}_c\right) \Vec{m} + \rho_{AHE} \Vec{m} \times \Vec{J}_c
\label{Main_Transport}
\end{equation}
where $\rho_0$ is the resistivity, $\rho_{AMR}$ and $\rho_{AHE}$ represent the coefficients for the anisotropic magnetoresistance and the anomalous Hall effect, respectively. $\Vec{J}_c$ is a charge current density that can flow in the sample on purpose or due to inductive and capacitive effects due to the coupling with the antenna and $\Vec{m}$ a unit vector pointing the magnetization direction. In the following we will consider that the antenna is fed by a continuous RF wave at the angular frequency $\omega$, which will induce magnetization dynamics around its equilibrium position at the same frequency as well as RF currents inside the sample, also at $\omega$ on which we can superimpose a DC offset:
\begin{align}\label{eq:m_Jc}
    \Vec{m} & =\vec{m}_{eq} + \Vec{\delta m}(\omega
    ) \\
    \Vec{J}_c & =  \Vec{J}_{c_0} + \Vec{J}_c(\omega)
\end{align}
Injecting this time variation (Eqs. \ref{eq:m_Jc} and 6) into Eq. \ref{Main_Transport}, it gives the full frequency contribution to the electric field (see Appendix \ref{Ap-A} Eq. \ref{E_y_full}-A5). The electric field $\vec{E}$ contains DC, $\omega$, $2\omega$, and $3\omega$ components. In rectified voltage measurements, only the DC and $2\omega$ terms contribute.

\subsection{Spurious effects}
The spurious effects in DC signal spin-pumping measurements only come from the induced currents. In this case $\Vec{J}_{c_0}=\vec{0}$. In our geometry, $\Vec{m}_{eq}=\vec{x}$ and the voltage is measured along $y$. Then, keeping only the DC (and $2\omega$) terms, the electric field simplifies to the following terms:
\begin{align}
   E_y = J^x_c(\omega) \left[ \rho_{amr} \delta m_y(\omega) + \rho_{ahe} \delta m_z(\omega)  \right]\label{Eq_EyH+}
\end{align}

Note that in Eq. \ref{Eq_EyH+} the first term change sign for $H<0$, i.e; $\Vec{m}_{eq}=\vec{x}$. From the linear response theory:
\begin{align}
   \delta m_y(\omega) & = \underline{\chi}_{yy} h_y(\omega) + \underline{\chi}_{yz} h_z(\omega)\\
   \delta m_z(\omega) & = \underline{\chi}_{zy} h_y(\omega) + \underline{\chi}_{zz} h_z(\omega)
\end{align}

Then : 
\begin{align}
   E_y & = J^x_c(\omega) \left[ \left( \rho_{amr}  \underline{\chi}_{yy} + \rho_{ahe} \underline{\chi}_{zy} \right) h_y(\omega) \right. \\
    & \left. + \left(\rho_{amr} \underline{\chi}_{yz} + \rho_{ahe} \underline{\chi}_{zz}\right) h_z(\omega) \right]
\end{align}\label{EqEy(w)}

Here, $ \underline{\chi}_{yy}$, $ \underline{\chi}_{zz}$, $ \underline{\chi}_{yz}$ and $ \underline{\chi}_{zy}$, represents the susceptibility coefficients in cartesian coordinates (See details in Appendix \ref{Ap-A}). Considering only the DC contribution and accounting for that the inductive and capacitive currents present a phase of $-\pi/2$ with respect to the RF magnetic field. The results obtained and valid for the two field regimes investigated in this study, $H>0$ and $H<0$ (see details in Appendix \ref{Ap-A}), are summarized below:

\begin{widetext}
\begin{align}
   \left<E_y\right> = -\frac{J^x_c}{2M_s} \left( \underbrace{\rho_{amr} \chi''_{yy}h_y + \rho_{ahe} \chi''_{zz}h_z}_{Symetric}+ \underbrace{\rho_{ahe} \chi''_{zy}h_y + \rho_{amr} \chi''_{yz}h_z}_{AntiSymetric} \right)  \text{for } H>0 \label{Eq_Ey_H+} \\
  \left<E_y\right> = -\frac{J^x_c}{2M_s} \left( \underbrace{-\rho_{amr} \chi''_{yy}h_y + \rho_{ahe} \chi''_{zz}h_z}_{Symetric}+ \underbrace{\rho_{ahe} \chi''_{zy}h_y + \rho_{amr} \chi''_{yz}h_z}_{AntiSymetric} \right)  \text{for } H<0 \label{Eq_Ey_H-}
\end{align} 
\end{widetext} 
 
For a magnetic layer with simple anisotropy, the susceptibility coefficient can be easily calculated. The important implication of the above result (Eq. 12 and \ref{Eq_Ey_H-}) is the shape of the Lorentzian components due to the different contributions of transversal AMR (PHE) and AHE. Eq. (\ref{Eq_Ey_H+}) is valid for a positive $H$-field sweep. The antisymmetric part is fully even with field reversal and the symmetric ones is even for the AHE contribution while is odd for the transversal AMR (PHE) contribution. These model explained our experimental results showed in Fig.\ref{Fig1}(b1-b3) for the different antennas, i.e $h_z$ in GAP, $h_y$ in SHORT, or a combination of both in FULL.

\subsection{Calibration of the RF field}

To calibrate the RF field, we inject a DC current and monitor the resulting evolution of the rectified amplitude voltage. This procedure requires compute the DC component of the transverse electric field, $E_y$. By retaining only the terms proportional to the $\mathbf{J}_{c,0}$, in Eq. (\ref{E_y_full})—as derived in Appendix A—the expression simplifies to:

\begin{align}
   \left< E_y \right> = \rho_0 J^y_{c_0} + \rho_{amr} J^y_{c_0} \left< \delta m_y^2(\omega) \right>
\end{align}

When expressing $\left< \delta m_y^2(\omega) \right>$ in the linear magnetization dynamics framework, it turns out that:
\begin{equation}
<\delta m^2_y(t)> = \frac{1}{2} |\underline{\chi}_{yy}|^2 h_y^2 + \frac{1}{2} |\underline{\chi}_{yz}|^2 h_z^2. 
\end{equation}
 $<\delta m^2_y(t)>$ is then the sum of the modules of two Lorentzian, which implies that it is a Lorentzian function symmetric around the resonance. 
 From this result, we derive the general expression at resonance (further analytical details are provided in Appendix \ref{Ap-A}):

\begin{align}\label{eq:m_dm}
     <\Vec{m} \times \frac{d\Vec{m}}{dt}> &= 2 \omega \frac{\chi'_{yz}}{\chi''_{yy}} <\delta m^2_y(t)> 
\end{align}

Here, $\langle \delta m_y^2 \rangle$ can be extracted from AMR measurements. Recall that when a DC current is injected along the $y$-direction in the spin-pumping bar, the rectified voltage reads
$V_y(t) = [R_0 + \Delta R_{\mathrm{AMR}}, \delta m_y^2(t)] I_{\mathrm{DC}}$.
The DC voltage variation induced by $I_{\mathrm{DC}}$ arises from time averaging,
$V_y^{\mathrm{DC}} = \langle V_y(t) \rangle = [R_0 + \Delta R_{\mathrm{AMR}} \langle \delta m_y^2(t) \rangle] I_{\mathrm{DC}}$.
This leads to the expression for the spin current, valid for all RF-field geometries considered in this work, namely along $\hat{y}$ and $\hat{z}$:

\begin{align}
    <\mathbf{J}_S^{\Vec{n}}>=\frac{\hbar}{4\pi}\frac{g^{\uparrow\downarrow}_{eff}}{M_s^2} 2 \omega \sqrt{\frac{H}{H+M_{eff}}} \frac{\Delta V_y^{DC} }{\Delta R_{AMR} I_{DC}} \Vec{m}_{eq} \label{eqJsn}
\end{align}

Thus, we derive an analytical expression that permits the full experimental determination of the spin current density $<\mathbf{J}_S^{\Vec{n}}>$ propagating along the $z$ direction, with its spin polarization oriented along $\Vec{m}_{eq}$. Equation~\ref{eqJsn} is equivalent to the effective spin current given in Eq.~\ref{eq:Js}, from which the relation for $h_{\mathrm{RF}}$ in Eq.~\ref{eqhrf} follows. Moreover, for in-plane magnetization under an in-plane DC field and neglecting in-plane anisotropy, the resonance condition reads
$\omega = \gamma_0 \sqrt{H_r (H_r + M_{\mathrm{eff}})}$,
or equivalently the resonance field
$H_r = \tfrac{1}{2}\left[-M_{\mathrm{eff}} + \sqrt{M_{\mathrm{eff}}^2 + 4(\omega_0/\gamma_0)^2}\right]$.
Under these conditions, Eq.~\ref{eqJsn} simplifies to:

\begin{equation}
    \mathbf{J}_S^{\Vec{n}}=\frac{\hbar}{4\pi}g^{\uparrow\downarrow}_{eff}2 \gamma_0 H_r \frac{\Delta V_x^{DC} }{\Delta R_{AMR} I_{DC}} \Vec{x} \label{EqJsn2}
\end{equation}

\section{Experimental protocol and example on single NiFe and NiFe/Pt bilayer}
\label{Section3}
\begin{figure}[h]
\centering
\includegraphics[width=0.5\textwidth]{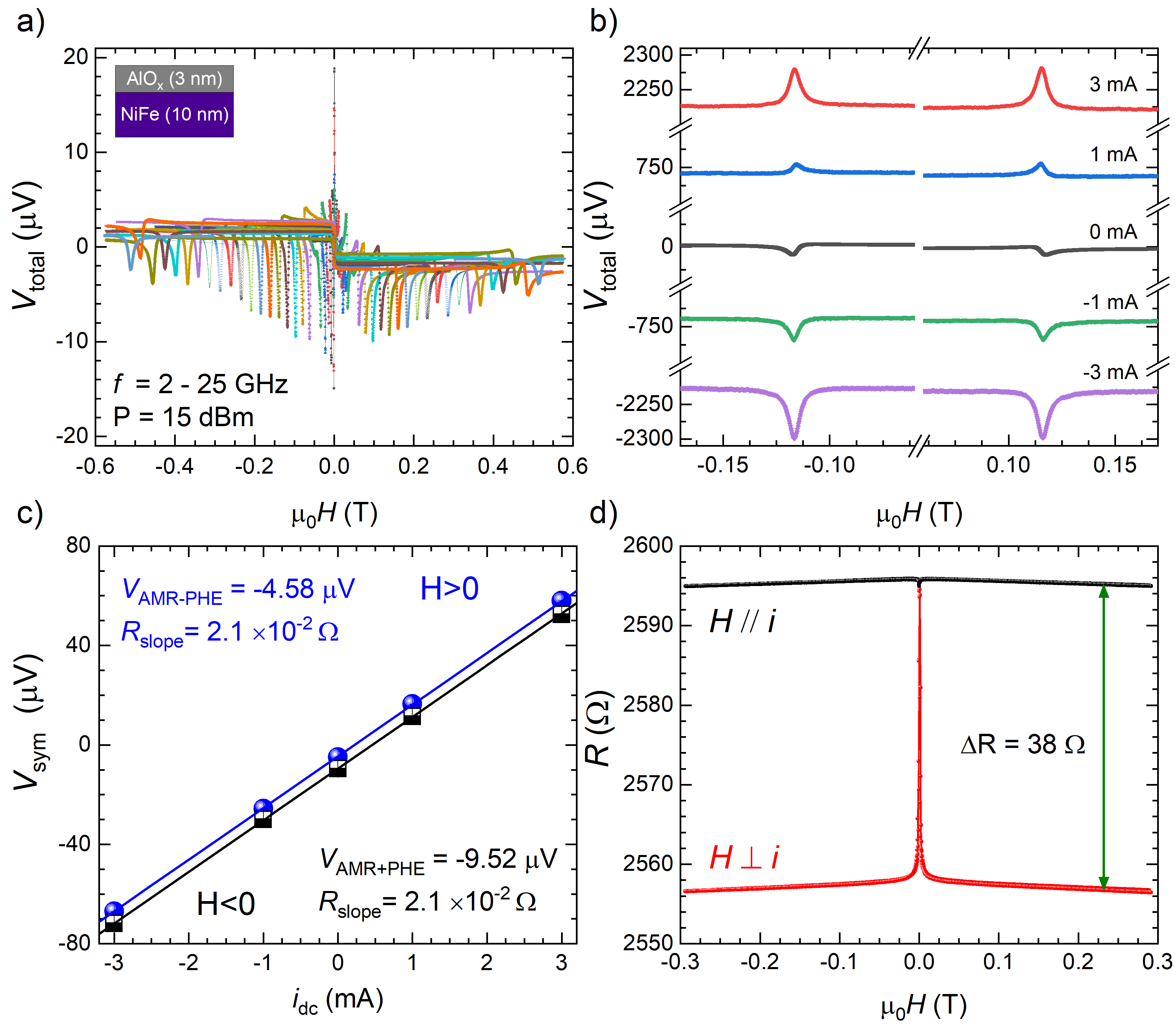}
\begin{quote}
\caption{
Experimental protocol to quantify $h_{\mathrm{RF}}$. (a) $V(H)$ sweeps at different microwave frequencies. (b) $V(H)$ response under varying DC bias currents, $i_{DC}$, at 10 GHz. (c) Extraction of the slope $\Delta V_{sym} / \Delta I_{DC}$ from the linear dependence of the symmetric amplitude on $i_{DC}$. (d) In-plane magnetoresistance measurements to obtain $\Delta R$ in parallel and perpendicular configurations. Combining these steps allows for the experimental determination of $h_{\mathrm{RF}}$, resulting in 0.5 G for the FULL geometry at 10 GHz.}\label{Fig_Protocol}
\end{quote}
\end{figure}

 To address the experimental determination of $h_{\mathrm{RF}}$, we develop the protocol described below. Although it requires more measurements than a single SP-FMR scan, it enables an accurate quantification of $h_{\mathrm{RF}}$ and, consequently, of the spin-to-charge conversion efficiency. Combinig Eq 2 and 17, $h_{\mathrm{RF}}$ can be calculated using the following expression for all antenna geometries:


{\footnotesize
\begin{equation}\label{eqhrf}
\mu_0h_{RF}= \alpha \mu_0(2H_{r}+M_{eff}) \sqrt\frac{H_{r}}{H_{r}+M_{eff}} \sqrt{\frac{\Delta V_{sym} }{\Delta  i_{DC}} \frac{2}{\Delta R_{AMR}}}
\end{equation}
}
\noindent where $\Delta R_{\mathrm{AMR}}$ denotes the change in longitudinal resistance due to AMR, while $\Delta V_{\mathrm{sym}}$ is the variation of the symmetric voltage amplitude under FMR conditions induced by the additional injection of $i_{\mathrm{DC}}$ into the SP-FMR or ST-FMR slabs. This setup-dependent parameter is crucial for accurately quantifying spin–charge interconversion. The experimental protocol is summarized in Fig.~\ref{Fig_Protocol} for a single NiFe (10 nm) film as follows:

\begin{enumerate}
    \item Measure the voltage $V(H)$ as a function of the applied DC magnetic field at fixed microwave power and frequency. By repeating the measurements at different microwave frequencies, we extract the damping $\alpha$ and the effective magnetization $M_{eff}$. 
       
    \item For a given frequency, apply a DC current $i_{DC}$ to the spin pumping bar of the sample. Measure $V(H)$ and repeat for different $i_{DC}$ values. The symmetric component will change linearly with the added DC bias current due to the AMR voltage, which is even with the $H$ field. This is illustrated in Fig. \ref{Fig_Protocol}b for a 10 GHz MW frequency.

    \item Determine $\frac{\Delta V_{sym} }{\Delta  i_{DC}}$, labeled as $R_{slope}$ in the Fig. \ref{Fig_Protocol}c. 
    
    \item Repeat the previous two steps for different microwave frequencies.
    \item The AMR-induced resistance variation, $\Delta R_{AMR}$, is obtained by measuring the longitudinal resistance with magnetic fields applied in-plane perpendicular and parallel to the bar, as illustrated in Fig.~\ref{Fig_Protocol}d. Alternatively, it can be determined from the in-plane angular dependence of the resistance at a fixed magnetic field exceeding the saturation and resonance fields.
    
\end{enumerate}
  
\begin{figure}[h]
\centering
\includegraphics[width=0.45\textwidth]{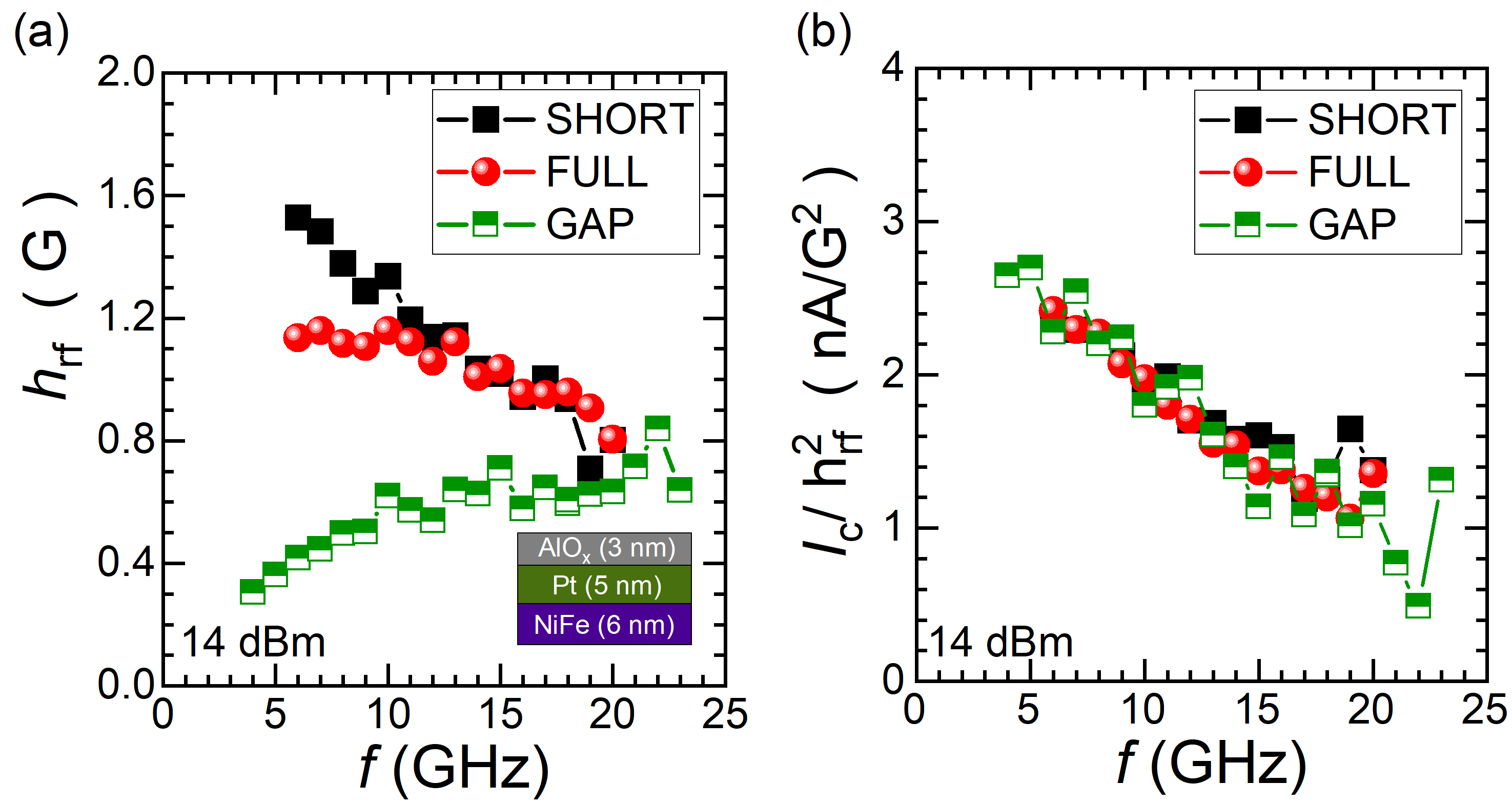}
\begin{quote}
\caption{(a) Strenght RF field $h_{\mathrm{RF}}$ for the three CPWs geometries on a //NiFe(6)/Pt(5) bilayer measured for different RF frequencies and calculated using Eq. \ref{eqhrf}.(b) Charge current production normalized by $h_{RF}^2$. We can see that the values are the same independently of the CPW geometry, validating the use of Eq. \ref{eqhrf} to estimate $h_{\mathrm{RF}}$ in the three different CPW.}\label{Fighrf_Py6Pt5}
\end{quote}
\end{figure}

 Fig. \ref{Fighrf_Py6Pt5}a shows the $h_{\mathrm{RF}}$ values as function of frequency obtained for the three types of antenna for a //NiFe(6)/Pt(5)/AlO$_x$(3) bilayer. If the measured voltage originates primarily from spin pumping and the subsequent ISHE in the Pt layer, the normalized charge current—defined as $I_c / h_{\mathrm{RF}}^2$ according to Eqs.~\ref{eq:Js} and \ref{eq:SPIc}—should be independent of the antenna configuration. This behavior is confirmed in Fig.~\ref{Fighrf_Py6Pt5}b, thereby validating both the theoretical model and the proposed experimental protocol. Further, we are also showing that he rectifying effects are negligible for 6 nm of NiFe.
 
\begin{figure}[h]
\centering
\includegraphics[width=0.45\textwidth]{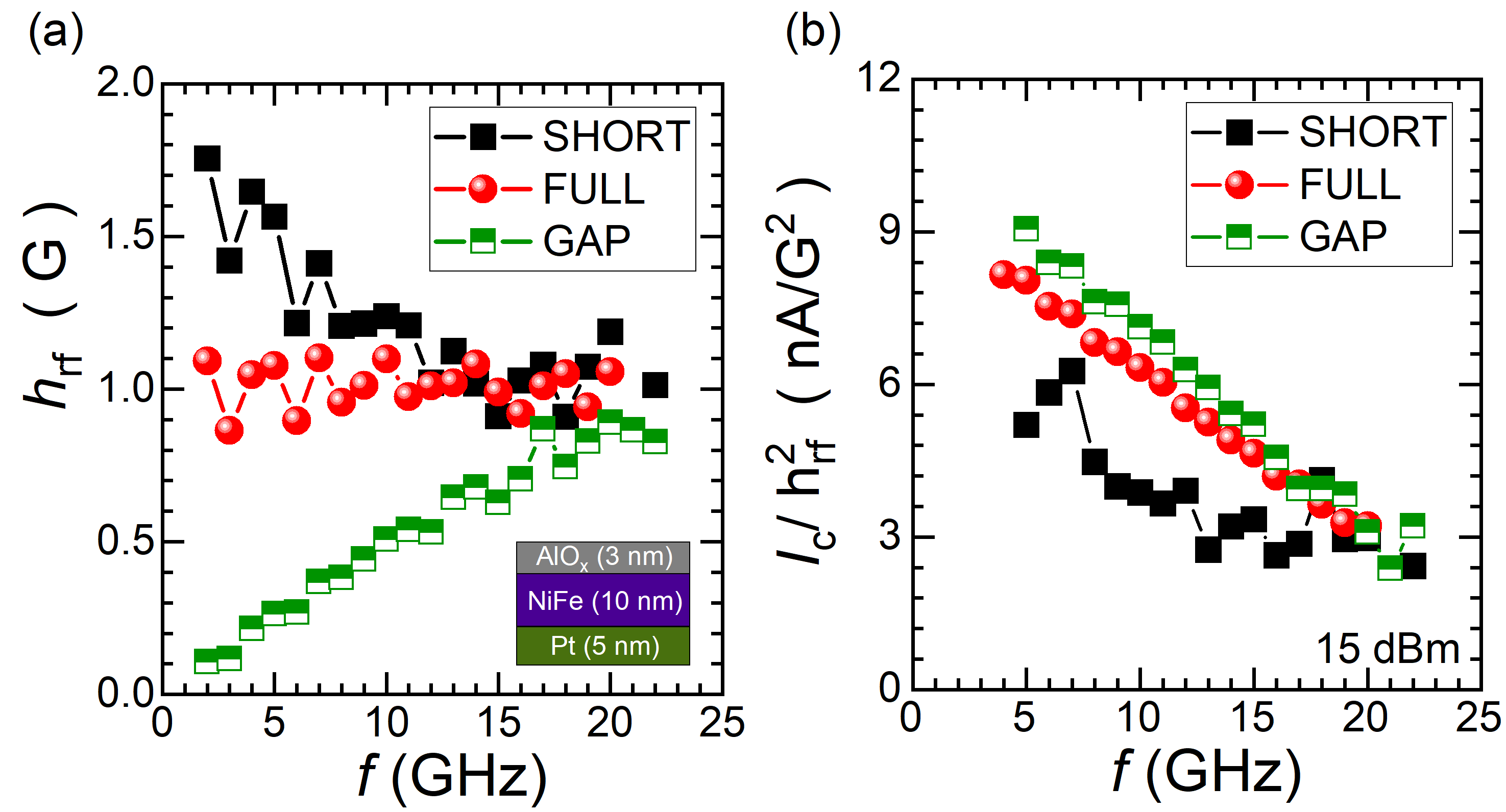}
\begin{quote}
\caption{(a) Extracted $h_{\mathrm{RF}}$ values for the three antenna geometries on a NiFe(10)/Pt(5) bilayer as a function of frequency, calculated using Eq. \ref{eqhrf}. (b) Charge current normalized by $h_{RF}^2$. The lack of overlap between the three geometries in (b) confirms the significant influence of the AMR ($V_{sym}^{odd}$) in the 10 nm NiFe layer, as shown in Fig. \ref{Fig1}.}\label{Fighrf_Pt5Py10}
\end{quote}
\end{figure}

After validating our protocol, we return to the //Pt/NiFe(10 nm) bilayer showed in Fig. \ref{Fig1}(c1-c3) to assess potential odd transversal AMR ($V_{sym}^{odd}$) contribution. Since ISHE-driven charge production is antenna geometry-independent, any variation in the $I_c/h_{RF}^2$ ratio among the antennas must stem from extrinsic effects like the transversal AMR ($V_{sym}^{odd}$). In Fig. \ref{Fighrf_Pt5Py10}a, the extracted $h_{\mathrm{RF}}$ values show a frequency evolution similar to our previous results. However,  the normalized current $I_c/h_{RF}^2$ (Fig. \ref{Fighrf_Pt5Py10}b) exhibits clear discrepancies between the three configurations. This divergence confirms that transversal AMR ($V_{sym}^{odd}$) contributions are significant at a NiFe thickness of 10 nm and may partially explain the wide dispersion of spin-to-charge conversion efficiencies reported in the literature.

\begin{figure}[h]
\centering
\includegraphics[width=0.45\textwidth]{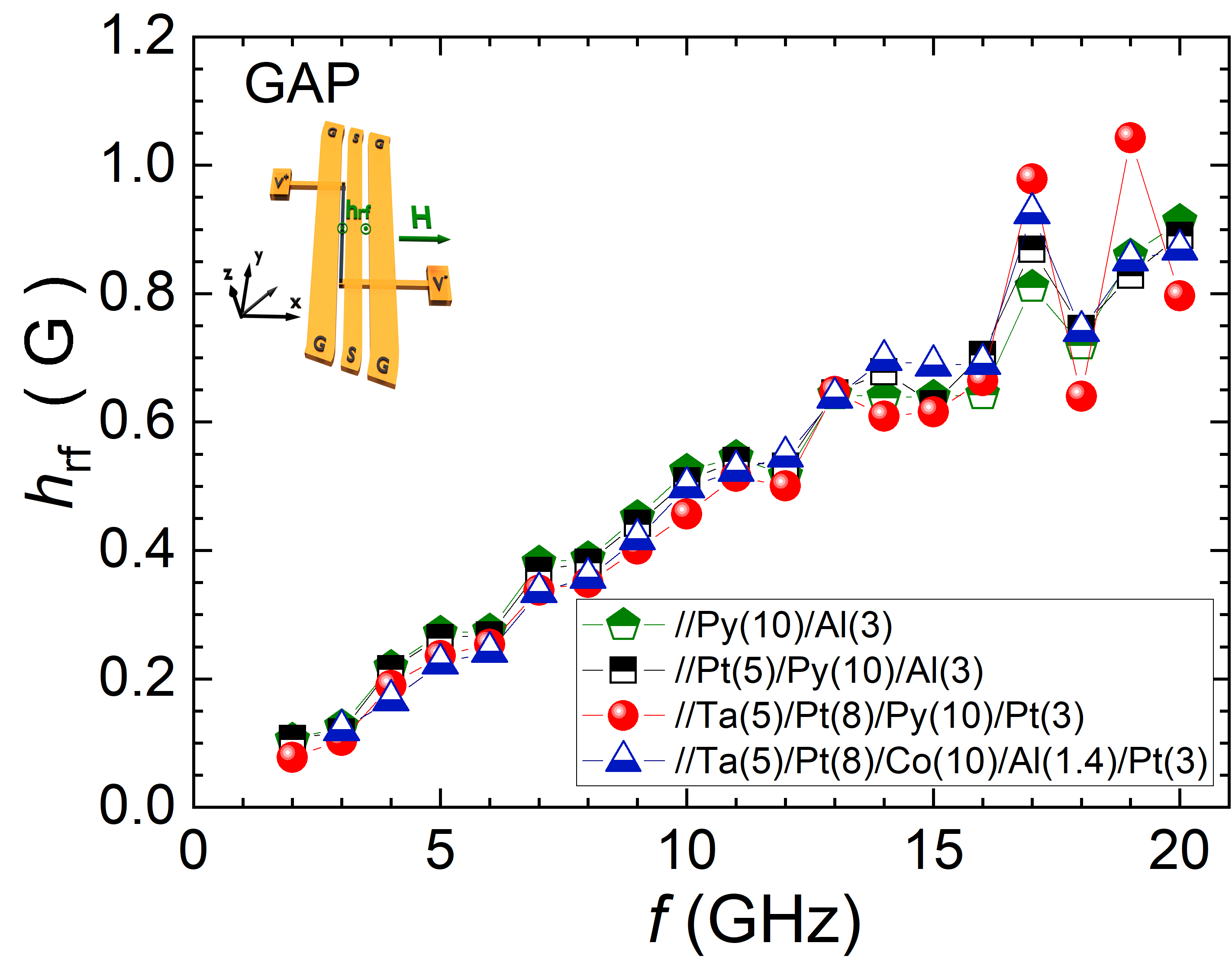}
\begin{quote}
\caption{The RF field strength $h_{\mathrm{RF}}$ for the GAP geometry is shown for several samples deposited on identical Si-SiO$2$ substrates. Utilizing Eq. \ref{eqhrf} with the measured of the total $\Delta R_{AMR}$ yields identical $h_{\mathrm{RF}}$ values across all samples, as expected. This agreement demonstrates that the determined field strength is a characteristic of the antenna geometry rather than the material stack.}\label{Fighrf_GAP}

\end{quote}
\end{figure}

It is essential to note that the AMR variation, $\Delta R_{\mathrm{AMR}}$, must correspond to the total value measured for the entire sample stack. While some works use the AMR of an isolated ferromagnetic layer \cite{Feng2012, Tshitoyan2015}, and others do not clearly specify this point \cite{Dong2023}, our results show that the total $\Delta R_{\mathrm{AMR}}$ of the bilayer or multilayer is the appropriate parameter in Eq.~\ref{eqhrf}.
To validate this, we measured $h_{\mathrm{RF}}$ for a wide range of materials—from single layers to complex multilayers—under identical input power and frequency. Since the antenna geometry and substrate were unchanged, $h_{\mathrm{RF}}$ is expected to be independent of the sample stack, which is indeed confirmed in Fig.~\ref{Fighrf_GAP}.
Specifically, we examined samples with markedly different transport properties: a NiFe (10 nm) single layer ($\Delta R_{\mathrm{AMR}}=38~\Omega$), a //Pt(5)/NiFe(10)/AlO$_x$ bilayer ($23~\Omega$), and a more conductive //Ta(5)/Pt(8)/Py(10)/Pt(3) multilayer ($12.4~\Omega$). Using the total measured AMR in Eq.~\ref{eqhrf} yields identical $h_{\mathrm{RF}}$ values in all cases. This result is further corroborated by a Co-based stack, //Ta(5)/Pt(8)/Co(10)/Al(1.4)/Pt(3) ($\Delta R_{\mathrm{AMR}}=10.4~\Omega$), which exhibits the same frequency dependence.
These results validate our protocol and ensure a robust determination of the RF field.

\section{Rectification signal in SPFMR and STFMR devices}
\label{Section4}
 Accurately quantifying spin-orbit parameters requires isolating rectification effects from spin pumping signals to avoid overestimating or underestimating values. Although the lineshape contributions from AMR and AHE have been extensively analyzed \cite{Azevedo2011, Rojas-Sanchez2013, Iguchi2017}, the odd symmetrical Lorentzian AMR component is often neglected by assuming a specific phase shift between the current and the magnetization \cite{Azevedo2011}. Our results show that both AMR and AHE contributions depend on the FM thickness and the direction of $h_{\mathrm{rf}}$ (Eq. 12 and \ref{Eq_Ey_H-}), in agreement with Ref.~\cite{Harder2016}. To further disentangle these effects, we compare experimental results obtained from 10 nm FM layers of Fe, Ni, NiFe, and CoFeB.

\begin{widetext}

\begin{figure}[h]
\centering
\includegraphics[width=1\textwidth]{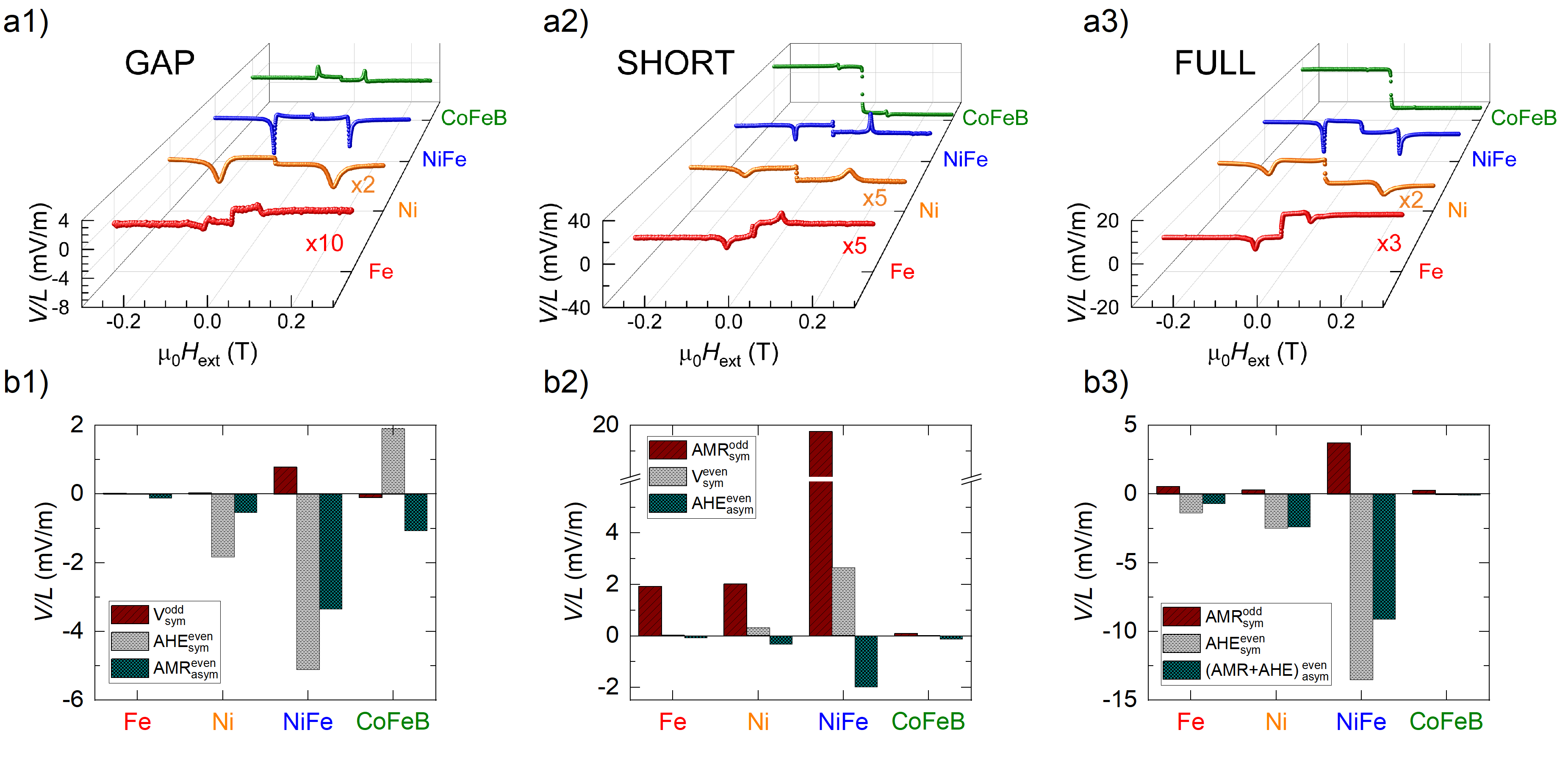}
\begin{quote}
\caption{Rectification effects in single FM layers measured on SP-FMR based devices. Voltage normalized by device length ($V/L$) versus external magnetic field at 10 GHz and 15 dBm MW input power for Si/SiO$_2$(500 nm)//FM(10 nm)/AlO$_x$(3 nm), FM = Fe, Ni, NiFe, CoFeB. a1–a3: raw signals; b1–b3: antisymmetrized values for GAP, SHORT, and FULL geometries. Fe and Ni signals are scaled for visibility. Focus in $V_{sym}^{odd}$, GAP (b1) shows the lowest values likely due to misalignment, SHORT (b2) exhibits the highest $V_{sym}^{odd}$ due to transversal AMR, also know as Planar Hall effect (PHE), and FULL (b3) is intermediate. Comparing FM materials in SHORT antenna, NiFe has the largest $V_{sym}^{odd}$ and CoFeB the smallest.}\label{Fig6}

\end{quote}
\end{figure}

\begin{table}[h]
\caption{Normalized voltage by the bar length for each antenna geometry, GAP ($L_G=520 \mu m$), SHORT ($L_S=70 \mu m$) and FULL ($L_F=576 \mu m$) for different FM and the corresponding extracted values for the Anisotropic Magnetoresistance (AMR) and the Anomalous Hall Effect (AHE) are included.}
\label{tab:my-table} 
\begin{tabular}{l|ccc|ccc|ccc|cc|}
\cline{2-12}
\multirow{2}{*}{}           & \multicolumn{3}{c|}{GAP}                                                 & \multicolumn{3}{c|}{SHORT}                                               & \multicolumn{3}{c|}{FULL}                                                & \multicolumn{1}{c|}{\multirow{2}{*}{$\Delta R_{AMR}/L_F$}} & \multicolumn{1}{l|}{\multirow{2}{*}{$\Delta R_{AHE}/W$}} \\ \cline{2-10}
                            & \multicolumn{1}{c|}{$V^{odd}_{sym}/L_G$ } & \multicolumn{1}{c|}{$V^{even}_{sym}/L_G$} & $V^{even}_{asym}/L_G$ & \multicolumn{1}{c|}{$V^{odd}_{sym}/L_S$ } & \multicolumn{1}{c|}{$V^{even}_{sym}/L_S$} & $V^{even}_{asym}/L_S$& \multicolumn{1}{c|}{$V^{odd}_{sym}/L_F$} & \multicolumn{1}{c|}{$V^{even}_{sym}/L_F$} & $V^{even}_{asym}/L_F$ & \multicolumn{1}{c|}{}                        & \multicolumn{1}{l|}{}                         \\ \cline{2-12} 
                            & \multicolumn{3}{c|}{$m V$/m}                                                 & \multicolumn{3}{c|}{$m V$/m}                                                 & \multicolumn{3}{c|}{$m V$/m}                                                 & \multicolumn{2}{c|}{$\Omega/\mu m$}                                                                     \\ \hline
\multicolumn{1}{|l|}{Fe}    & \multicolumn{1}{c|}{0.027}          & \multicolumn{1}{c|}{-0.005}        & -0.116      & \multicolumn{1}{c|}{1.929}          & \multicolumn{1}{c|}{0.021}        & -0.067      & \multicolumn{1}{c|}{0.560}          & \multicolumn{1}{c|}{-1.349}        & -0.672      & \multicolumn{1}{c|}{0.014}                       & 0.109                                             \\
\multicolumn{1}{|l|}{Ni}    & \multicolumn{1}{c|}{0.031}          & \multicolumn{1}{c|}{-1.825}        & -0.533      & \multicolumn{1}{c|}{2.021}          & \multicolumn{1}{c|}{0.305}        & -0.311      & \multicolumn{1}{c|}{0.312}          & \multicolumn{1}{c|}{-2.479}        & -2.379      & \multicolumn{1}{c|}{0.028}                       & -0.017                                             \\
\multicolumn{1}{|l|}{NiFe}  & \multicolumn{1}{c|}{0.783}          & \multicolumn{1}{c|}{-5.111}        & -3.351      & \multicolumn{1}{c|}{19.759}          & \multicolumn{1}{c|}{2.660}        & -1.990      & \multicolumn{1}{c|}{3.720}          & \multicolumn{1}{c|}{-13.525}        & -9.111      & \multicolumn{1}{c|}{0.066}                       & 0.019                                             \\
\multicolumn{1}{|l|}{CoFeB} & \multicolumn{1}{c|}{-0.105}          & \multicolumn{1}{c|}{1.915}        & -1.068      & \multicolumn{1}{c|}{0.095}          & \multicolumn{1}{c|}{-0.010}        & -0.122     & \multicolumn{1}{c|}{0.269}          & \multicolumn{1}{c|}{-0.023}        & -0.068     & \multicolumn{1}{c|}{0.024}                       & 0.215                                             \\ \hline
\end{tabular}
\end{table}

\end{widetext}

Figure~\ref{Fig6} shows the voltage normalized by the slab length ($V/L$) for three CPW geometries. In the GAP antenna, where $h_{\mathrm{rf}}$ is perpendicular to the strip, a dominant symmetric even Lorentzian from AHE is observed in NiFe, Ni, CoFeB, and Fe, followed by an even antisymmetric component from transversal AMR, according to our results in Eq. \ref{Eq_Ey_H+} and \ref{Eq_Ey_H-}. A small symmetric–odd contribution arises from unavoidable experimental misalignment between the sample bar and the GSG track, which are not exactly parallel. In this geometry, thermal effects are strongly suppressed, as evidenced by the reduced zero-field jump compared to other configurations.

 In the SHORT antenna geometry (Fig.~\ref{Fig6}a2), Fe, Ni, and NiFe exhibit symmetric Lorentzians that are odd under magnetic field reversal, $V_{sym}^{odd}$, characteristic of transversal AMR with h$_{RF}$ along $y$ direction. This signature is indistinguishable from the spin- or orbital-pumping voltage in FM/HM bilayers such as NiFe/Pt. As a result, for a positive spin Hall angle, $V_{sym}^{odd}$ adds constructively to the spin-pumping signal, leading to an overestimation of the spin–charge conversion efficiency, while in inverted stacks it can partially cancel the signal. As shown in Fig.~\ref{Fig6}b2, NiFe exhibits the largest $V_{sym}^{odd}$ contribution (up to 20 mV/m), followed by Ni (2 mV/m) and Fe (1.9 mV/m). The remaining symmetric even voltage is attributed to thermal contributions due to Anomalous Nernst and Rigied Leduc effects, which are enhanced in this configuration and relative large for NiFe, consistent with the pronounced zero-field jump and our thermal spurious effects model (Appendix~\ref{App:thermal}). The antisymmetric Lorentzian even component comes from the AHE. This is observed for Ni and NiFe.

In the FULL geometry, a dominant symmetric even signal is observed for NiFe, Ni, and Fe (Fig.~\ref{Fig6}a3), evidencing that AHE dominates in this configuration. This means that the contribution with $h_z$ are more significant than those with $h_y$. Overall, among the three geometries, the GAP antenna—with $h_{\mathrm{RF}}$ perpendicular to the sample—provides the most reliable conditions for spin-pumping experiments by strongly suppressing spurious signal due to transversal AMR ($V_{sym}^{odd}$) that hinder accurate spin–orbit quantification. Among all the metallic FMs presented in this work, CoFeB is the one that exhibits the least spurious effects.

Table~\ref{tab:my-table} summarizes the normalized voltage components—$V_{sym}^{odd}$, $V_{sym}^{even}$, and $V_{asym}^{even}$—divided by the respective bar lengths for the GAP, SHORT, and FULL antenna geometries across the four FM materials studied (Fe, Ni, NiFe, CoFeB). This normalization accounts for geometric scaling, enabling direct comparison of inductive contributions.
The table also reports transport coefficients: the anisotropic magnetoresistance ($\Delta R_{\mathrm{AMR}}$) measured on the same spin-pumping bars, and the anomalous Hall effect (AHE) measured using double Hall cross bars with a magnetic field applied  perpendicular to the film plane. $\Delta R_{AHE}$ measurements show that NiFe, Fe, and CoFeB exhibit anomalus Hall effect sign opposite to Ni, consistent with previous reports \cite{Omori2019,Zhang2013}. For instance, it has been shown that the AHE sign for NiFe depends on its layer thickness \cite{Omori2019,Zhang2013}. Direct quantitative comparison between transport-derived and spin-pumping–extracted AMR and AHE values is not straightforward, as the effective AHE voltage in spin pumping experiments also includes susceptibility terms defined in Eqs.~\ref{Eq_Ey_H+} and \ref{Eq_Ey_H-}, which depend on parameters such as saturation magnetization (Eq.~\ref{chi''}) and magnetic damping (Eqs.~\ref{omega0 and delta}).
Among these, the diagonal susceptibility components ($\chi''_{xx}$ and $\chi''_{yy}$) dominate over the off-diagonal term ($\chi''_{yz}$), as evidenced experimentally in SHORT (Fig.~\ref{Fig6}b2) and GAP (Fig.~\ref{Fig6}b1) geometries. In SHORT, the AMR signal dominates over AHE, whereas in GAP, AHE dominates over AMR. 
\begin{figure}[h]
\centering
\includegraphics[width=0.45\textwidth]{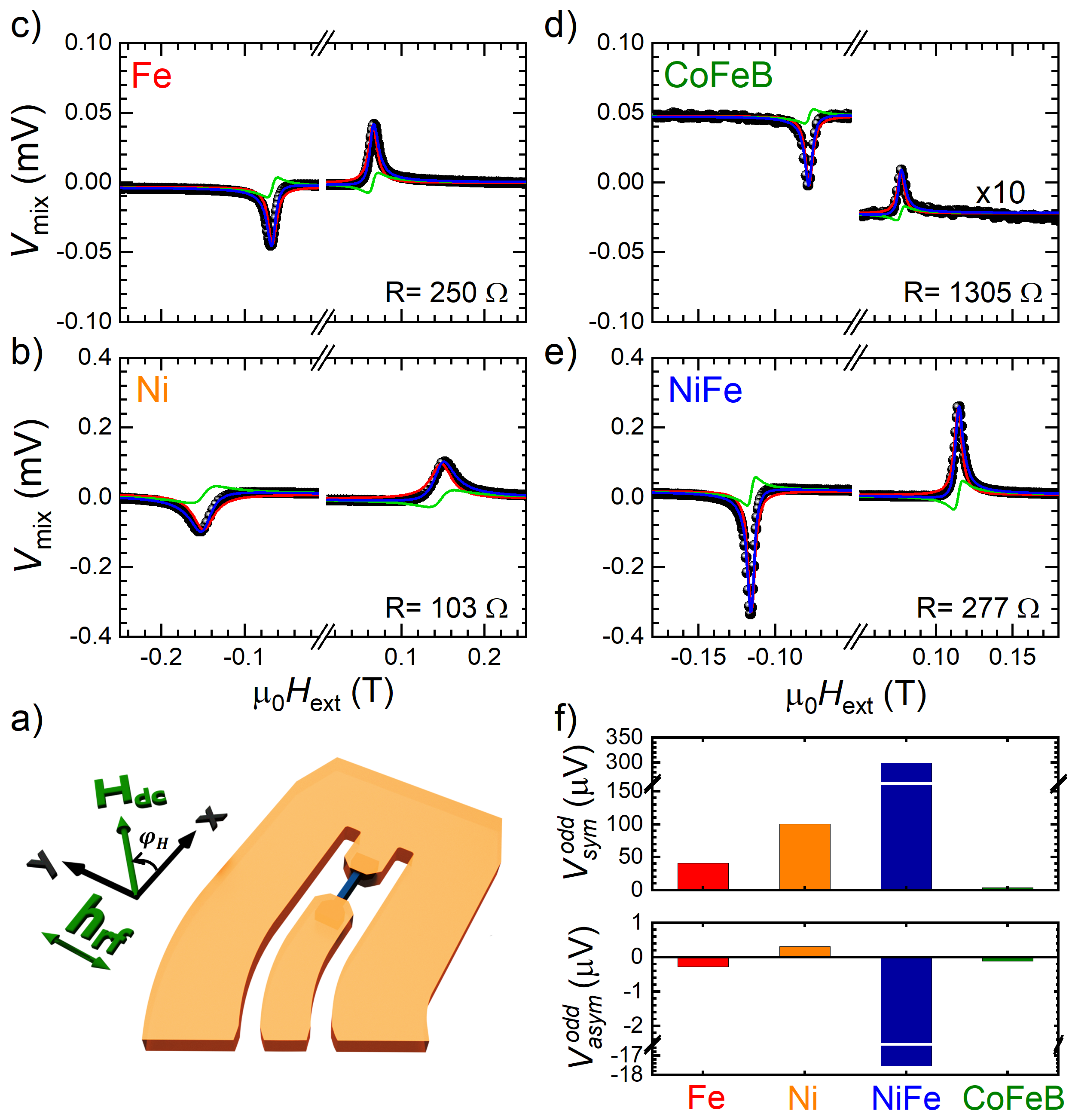}
\begin{quote}
\caption{a) Schematic of the ST-FMR device. The 10 $\mu$m $\times$ 60 $\mu$m sample is part of the antenna, with Ti/Au electrodes. A microwave current is injected in-plane, and the DC voltage is measured as a function of an external magnetic field applied at $\varphi_H = 45^\circ$. b–e) Rectified voltage versus field at 10 GHz and 15 dBm MW input power for Si/SiO$_2$(500 nm)//FM(10 nm)/Al(3 nm) layers, FM = Fe, Ni, NiFe, CoFeB (CoFeB signal $\times$10). All layers show a symmetric peak reversing sign with field (red fits), indicating AMR origin which mimic a SOT-like symmetry. The largest are for NiFe and Ni, smaller in Fe, and minimal in CoFeB.} \label{Fig7-v}
\end{quote}
\end{figure}

A typical ST-FMR device closely resembles the SHORT antenna geometry, which we identified as the least suitable for isolating pure spin-pumping signals. Therefore, evaluating rectification effects in single FM layers is crucial. The SP-FMR analytical expressions (Eqs.~\ref{Eq_Ey_H+} and \ref{Eq_Ey_H-}) cannot be directly applied to ST-FMR due to differences in device geometry and underlying mechanisms, as the magnetization equilibrium is typically set at 45° relative to the injected AC current along $x$ and $h_{\mathrm{RF}}$ along $y$ (Fig. ~\ref{Fig7-v}a) \cite{ gudin2023isotropic, Saglam2018, Guillemard2018Charge-spinDirections, Liu2019, CespedesBerrocal2021, Damas2022, Damas2025, damas2023}. Despite these differences, the rectified signal remains clearly identifiable. Figure~\ref{Fig7-v}b–e shows that all single FM layers exhibit a symmetric voltage peak reversing with the magnetic field, consistent with an AMR origin. This self-induced signal mimics the symmetry of the damping-like spin-orbit torque (SOT) in ST-FMR, as previously observed in Ta/Ni bilayers \cite{Liu2025}. The symmetric–odd voltage magnitudes from the SHORT antenna match the odd-symmetric components extracted via ST-FMR, peaking in NiFe (298 $\mu$V, 5 V/m), and minimizing in CoFeB (4 $\mu$V, 67 mV/m).

Our results also provide a methodological explanation for discrepancies in the literature, suggesting that much of the reported SOT efficiency enhancement in Ni-based heterostructures \cite{Lee2021} actually arises from material-specific rectification backgrounds \cite{Liu2025}.

\section{Conclusions}
In this work, we introduce a comprehensive model and experimental protocol designed to quantify RF field intensity across various antenna geometries.
The method involves applying a DC bias current to sample bars during voltage-versus-field measurements and has been validated across a range of systems, from single ferromagnetic layers to complex multilayers
A comparative analysis of 10 nm-thick ferromagnetic (FM) thin films reveals that rectification effects—arising from the AHE and transverse AMR (also known as PHE)—are minimal in Fe and CoFeB, whereas Ni and NiFe exhibit significantly enhanced signals. Notably, the symmetry of these AMR-rectified voltages closely mimics the lineshapes characteristic of spin-pumping or spin-transfer torque ferromagnetic resonance (ST-FMR) when the RF magnetic field ($h_{\mathrm{RF}}$) is oriented in-plane. Specifically, these signals manifest as a symmetric Lorentzian centered at the resonance field ($H_{\mathrm{res}}$) that is odd under reversal of the DC magnetic field ($H_{\mathrm{DC}}$), potentially complicating the isolation of pure spin-current contributions.
These observations underscore the importance of accounting for rectification effects to ensure the reliable characterization of spin-to-charge interconversion.
For example, rectification effects are found to be negligible in the NiFe(6)/Pt(5) bilayer, in contrast to NiFe(10)/Pt(5). 
By disentangling pure spin-current signals from rectification artifacts, our protocol provides a pathway to resolve discrepancies in the reported spin-charge current interconversion efficiencies across diverse spin-orbit materials in the literature. Importantly, this protocol and the associated rectification corrections are equally applicable to systems where orbital angular momentum contributions are present.
Our findings not only offer guidelines for optimizing future heterostructures but also hold significant implications for both fundamental research and practical applications.

\medskip
\textbf{Acknowledgements}\\ 
This work was funded by the ERC CoG project MAGNETALLIEN grant ID  101086807, the EU-H2020-RISE project Ultra Thin Magneto Thermal Sensoring ULTIMATE-I (Grant ID. 101007825), and the French National Research Agency (ANR) through the project “Lorraine Université d’Excellence” reference ANR-15-IDEX-04-LUE. It was also partially supported by the ANR through the France 2030 government grants EMCOM (ANR-22-PEEL-0009), PEPR SPIN ANR-22-EXSP-0007 and ANR-22-EXSP-0009. Devices in the present study were patterned at Institut Jean Lamour's clean room facilities (MiNaLor). M. Yactayo is grateful to MSCA RISE ULTIMATE-I (Grant ID. 101007825) and CONCYTEC for the funding provided through the PROCIENCIA under Undergraduate and Postgraduate Theses in Science, Technology and Technological Innovation 2025-01 program (PE501098380-2025)

\bibliography{biblioRojas_2025}

\appendix

\clearpage 

\section{Supplementary}
\label{Ap-A}
\renewcommand{\thesubsection}{\thesection\arabic{subsection}}
\subsection{Full expression of the electric field $E_y(w)$}

By substituting the time-variation expressions from Eqs. (\ref{eq:m_Jc}) and (6) into the Eq. (\ref{Main_Transport}), the complete expression for the rectified electric field is obtained as:


\begin{widetext}
\begin{align}
   \Vec{E} = & \rho_0 \Vec{J}_{c_0} + \rho_{amr} \left(\Vec{m}_{eq}.\Vec{J}_{c_0} \right) \Vec{m}_{eq} + \rho_{ahe} \Vec{m}_{eq} \times \Vec{J}_{c_0} \\
        + &\rho_0 \Vec{J}_c(\omega) + \rho_{amr} \left(\Vec{m}_{eq}.\Vec{J}_c(\omega) \right) \Vec{m}_{eq} + \rho_{ahe} \Vec{m}_{eq} \times \Vec{J}_c(\omega) \\
        + &\rho_{amr} \left( \Vec{\delta m}(\omega).\Vec{J}_{c_0} \right) \Vec{m}_{eq} + \rho_{amr} \left( \Vec{m}_{eq}.\Vec{J}_{c_0} \right) \Vec{\delta m}(\omega) + \rho_{ahe} \Vec{\delta m}(\omega) \times \Vec{J}_{c_0} \\
        + & \rho_{amr} \left( \Vec{\delta m}(\omega).\Vec{J}_c(\omega) \right) \Vec{m}_{eq} + \rho_{amr} \left( \Vec{m}_{eq}.\Vec{J}_c(\omega) \right) \Vec{\delta m}(\omega) + \rho_{amr} \left( \Vec{\delta m}(\omega).\Vec{J}_{c_0} \right) \Vec{\delta m}(\omega) + \rho_{ahe} \Vec{\delta m}(\omega) \times \Vec{J}_c(\omega)\\
        + & \rho_{amr} \left( \Vec{\delta m}(\omega).\Vec{J}_c(\omega) \right) \Vec{\delta m}(\omega)
\end{align}\label{E_y_full}
\end{widetext}

\subsection{Properties of the susceptibility matrix}

Considering the magnetization free energy $\epsilon (\theta , \varphi)$ in spherical coordinates. Noting $\epsilon_\theta$ and $\epsilon_\varphi$ the first derivatives respectively with $\theta$ and $\varphi$ and $\epsilon_{\theta\theta}$, $\epsilon_{\varphi\varphi}$ and $\epsilon_{\theta\varphi}=\epsilon_{\varphi\theta}$ (the egality is due to the Schwartz theorem) the second derivatives $(\theta_0,\varphi_0)$ denotes the equilibrium position.\\

Considering a perturbative harmonic RF field in the complex formalism $\uline{\delta \vec{h}}(\omega)=\uline{\delta \vec{h}} e^{i \omega t}$ that will induced the magnetization dynamics $(\delta\theta,\delta\varphi)$ around its equilibrium position. In the linear approximation of the magnetization dynamics in the transverse to the equilibrium position $ \uline{\delta \vec{m}}$, the two quantities are link by the complex susceptibility matrix $ \uuline{\chi}(\omega)$ such that:
\begin{equation}
    M_{S}  \uline{\delta \vec{m}} (\omega)= \uuline{\chi} (\omega) . \uline{\delta \vec{h}}(\omega)=\left( \uuline{\chi}'+ i \uuline{\chi}''\right)\uline{\delta \vec{h}}  e^{i \omega t} 
    \label{susceptibility}
\end{equation}

one can express the real and imaginary part of the susceptibility matrix in term of the second derivative of the free magnetic density energy evaluated in  $(\theta_0,\varphi_0)$ \cite{damas2023}:

\begin{widetext}
\begin{subequations}
\begin{align}
    & \uuline{\chi}'(\omega) = \dfrac{\gamma_0^2 M_S}{(1+\alpha^2)\left[(\omega_0^2-\omega^2)^2 +(\Delta_\omega \omega)^2\right]} 
    \begin{bmatrix} 
    \dfrac{\varepsilon_{\varphi\varphi}(\omega_0^2 - \omega^2)}{\mu_0 M_S\sin^2{(\theta_0)}} + \dfrac{\alpha \Delta_\omega \omega^2 }{\gamma_0} & -\dfrac{\varepsilon_{\theta\varphi}(\omega_0^2-\omega^2)}{\mu_0 M_S \sin{(\theta_0})} + \dfrac{\Delta_\omega \omega^2}{\gamma_0} \\
     \noalign{\vspace{6pt}} 
    -\dfrac{\varepsilon_{\theta\varphi}(\omega_0^2-\omega^2)}{\mu_0 M_S \sin{(\theta_0)}}-\dfrac{\Delta_\omega \omega^2}{\gamma_0} & \dfrac{\varepsilon_{\theta\theta}(\omega_0^2-\omega^2)}{\mu_0 M_S}  + \dfrac{\alpha \Delta_\omega \omega^2}{\gamma_0} \end{bmatrix} \label{1chi'} \\
\text{and} \nonumber &  \\
    & \uuline{\chi}''(\omega) = \dfrac{\gamma_0^2 M_S \omega}{(1+\alpha^2)\left[(\omega_0^2 - \omega^2)^2 + (\Delta_\omega \omega)^2 \right]} 
    \begin{bmatrix} 
    -\dfrac{\Delta_\omega \varepsilon_{\varphi\varphi}}{\mu_0 M_S \sin^2{(\theta_0)}} + \dfrac{\alpha(\omega_0^2-\omega^2)}{\gamma_0} & \dfrac{\omega_0^2-\omega^2}{\gamma_0}+\dfrac{\Delta_\omega \varepsilon_{\theta\varphi}}{\mu_0 M_S\sin{(\theta_0)}} \\
     \noalign{\vspace{6pt}} 
     -\dfrac{\omega_0^2 - \omega^2}{\gamma_0} + \dfrac{\Delta_\omega \varepsilon_{\theta\varphi}}{\mu_0 M_S \sin{(\theta_0)}} & -\dfrac{\Delta_\omega \varepsilon_{\theta\theta}}{\mu_0 M_S} + \dfrac{\alpha(\omega_0^2 - \omega^2)  }{\gamma_0}
     \end{bmatrix} \label{1chi''}
\end{align}
\end{subequations}
\end{widetext}

where we have defined the resonance frequency $\omega_0$ and the linewidth of the resonance peak $\Delta_\omega$ as follows:
\begin{align}
    \omega_0 &= \frac{\gamma_0}{\sqrt{1+\alpha^2} \mu_0 M_S \sin{(\theta_0)}} \sqrt{\varepsilon_{\theta\theta} \varepsilon_{\varphi\varphi} -(\varepsilon_{\theta\varphi})^2} \\ 
    \Delta_\omega &= \frac{\gamma_0 \alpha}{(1+\alpha^2) \mu_0 M_S}\left(\varepsilon_{\theta\theta} + \frac{\varepsilon_{\varphi\varphi}}{\sin^2{(\theta_0)}} \right)
    \label{omega0 and delta}
\end{align}

We now examine the properties of the susceptibility matrix coefficients. 
In most relevant scenarios $\epsilon_{\theta\varphi}(\theta_0,\varphi_0)=0 $. Moreover, if we consider a material in the low-damping limit ($\alpha \ll 1$), retaining only terms to first order in $\alpha$, the susceptibility matrices reduce to:

\begin{widetext}
\begin{subequations}
\begin{align}
    & \uuline{\chi}'(\omega) = \dfrac{\gamma_0^2 M_S}{(\omega_0^2-\omega^2)^2 +(\Delta_\omega \omega)^2} 
    \begin{bmatrix} 
    \dfrac{\varepsilon_{\varphi\varphi}(\omega_0^2 - \omega^2)}{\mu_0 M_S\sin^2{(\theta_0)}} & \dfrac{\Delta_\omega \omega^2}{\gamma_0} \\
     \noalign{\vspace{6pt}} 
    -\dfrac{\Delta_\omega \omega^2}{\gamma_0} & \dfrac{\varepsilon_{\theta\theta}(\omega_0^2-\omega^2)}{\mu_0 M_S} \end{bmatrix} \label{chi'} \\
\text{and} \nonumber &  \\
    & \uuline{\chi}''(\omega) = \dfrac{\gamma_0^2 M_S \omega}{(\omega_0^2 - \omega^2)^2 + (\Delta_\omega \omega)^2} 
    \begin{bmatrix} 
    -\dfrac{\Delta_\omega \varepsilon_{\varphi\varphi}}{\mu_0 M_S \sin^2{(\theta_0)}} + \dfrac{\alpha(\omega_0^2-\omega^2)}{\gamma_0} & \dfrac{\omega_0^2-\omega^2}{\gamma_0} \\
     \noalign{\vspace{6pt}} 
     -\dfrac{\omega_0^2 - \omega^2}{\gamma_0} & -\dfrac{\Delta_\omega \varepsilon_{\theta\theta}}{\mu_0 M_S} + \dfrac{\alpha(\omega_0^2 - \omega^2)  }{\gamma_0}
     \end{bmatrix} \label{chi''}
\end{align}
\end{subequations}
\end{widetext}

Within the framework of this hypothesis, one obtains the following relations between the coefficients:
\begin{align}
    \chi'_{\theta\varphi} = - \chi'_{\varphi\theta} ~~ \text{and} ~~  \chi''_{\theta\varphi} = - \chi''_{\varphi\theta}
\end{align}

From Eqs. (A10), we can identify that the diagonal components of $\uuline{\chi}'$ are antisymmetric Lorentzian functions, whereas the off-diagonal components are symmetric. In contrast, for $\uuline{\chi}''$, the diagonal components are symmetric and the off-diagonal components are antisymmetric (vanish at the resonance condition).\\

At resonance, i.e., when $\omega = \omega_0$, the following straightforward simplifications appear:
\begin{align}\label{eq:sime_sucep}
    \chi'_{\theta\theta} = \chi'_{\varphi\varphi} = 0 ~~ \text{and} ~~  \chi''_{\theta\varphi} =  \chi''_{\varphi\theta}=0
\end{align}

There is also a less straightforward relation:
\begin{align}\label{eq:sucep_yz}
   {\chi'_{\theta\varphi}}^2 = \chi''_{\theta\theta}\chi''_{\varphi\varphi}
\end{align} 

It is relatively straightforward to translate the susceptibility coefficients from spherical to cartesian coordinates. In the case of interest in this paper, where the equilibrium position lies along $\vec{x}$:

\begin{align*}
    \chi_{yy} & = \chi_{\varphi\varphi}(\pi/2,0) \\
    \chi_{zz} & = \chi_{\theta\theta}(\pi/2,0) \\
    \chi_{yz} & = -\chi_{\varphi\theta}(\pi/2,0) \\
\end{align*}

\subsection{Demonstration of the Equation \ref{eq:m_dm}}
To derive Eq. \ref{eq:m_dm}, which we recall here for clarity:
\begin{align}
     \left< \Vec{m} \times \frac{d\Vec{m}}{dt} \right> &= 2 \omega \frac{\chi'_{yz}}{\chi''_{yy}} <\delta m^2_y(t)> \vec{x}
\end{align}

We first express the expression, $ \left< \Vec{m} \times \frac{d\Vec{m}}{dt} \right>$, in terms of the susceptibility matrix elements. Assuming the equilibrium magnetization is aligned along the $x$-axis ($\vec{m}_{eq}=m_x$) and considering and small variation around it, the dynamic components are given by:

\begin{align}
    \left< \Vec{m} \times \frac{d\Vec{m}}{dt} \right> =\left< \delta m_y \dot{\delta m}_z - \delta m_z \dot{\delta m}_y \right> \vec{x}
\end{align}
When dealing with product of linear terms, it is necessary to give up on the the linear formalism. Then $\vec{h}(\omega)=\vec{h} e^{i \omega t} \rightarrow \vec{h}(\omega)=\vec{h} \cos(\omega t)$ :
\begin{align}\label{eq:dmt}
   \delta m_j(t) & = \frac{1}{M_s} \sum_{k \in \{y,z\}} \left( \chi'_{jk} \cos{(\omega t)} - \chi''_{jk} \sin{(\omega t)} \right) h_k \\
   \dot{\delta m_j}(t) & = -\frac{\omega}{M_s} \sum_{k \in \{y,z\}} \left( \chi'_{jk} \sin{(\omega t)} + \chi''_{jk} \cos{(\omega t)} \right) h_k 
\end{align}
with $j \in \{y,z\}$ 

Then the product give when kipping only the DC terms:
\begin{align*}
   \left< \delta m_j(t) \delta\dot{m}_i(t) \right> = \frac{\omega}{2M^2_s} \sum_{k,l \in \{y,z\}} \left( \chi''_{jk}\chi'_{il} - \chi'_{jk}\chi''_{il}  \right) h_kh_l
\end{align*}
Which give: 
\begin{align}\label{m_dmt_full}
    \left< \Vec{m} \times \frac{d\Vec{m}}{dt} \right> & = \frac{\omega}{M^2_s} \left(
    \left(\chi'_{yz}\chi''_{yy} -\chi'_{yy}\chi''_{yz} \right) h_y^2 \right.\\
        & +  \left( \chi'_{yz}\chi''_{zz} - \chi'_{zz}\chi''_{yz}  \right) h_z^2 \\
        & \left. +\left(\chi'_{yy}\chi''_{zz} - \chi'_{zz}\chi''_{yy}  \right) h_yh_z  \right) \vec{x}
\end{align}

At the resonant condition $\omega = \omega_0$, the susceptibility relations from Eq. \ref{eq:sime_sucep} allow Eq. \ref{m_dmt_full}-A20 to be reduced to:

\begin{align}\label{eq:m_dmt_2}
    \left< \Vec{m} \times \frac{d\Vec{m}}{dt} \right> & = \frac{\omega}{M^2_s} \chi'_{yz} \left( \chi''_{yy} h_y^2 + \chi''_{zz} h_z^2  \right) \vec{x}
\end{align}

Considering the term $<\delta m^2_y(t)>$ from Eq. \eqref{eq:dmt} and retaining only the DC components, we obtain:

\begin{align}
    <\delta m^2_y(t)> & =  \frac{1}{2M^2_s} \sum_{k,l \in \{y,z\}} \left( \chi'_{yk}\chi'_{yl} + \chi''_{yk}\chi''_{yl} \right) h_kh_l \\
                    & =  \frac{1}{2M^2_s} \left( \left( \chi'_{yy}\chi'_{yy} + \chi''_{yy}\chi''_{yy} \right) h_y^2 \right. \\
                    & + \left. \left( \chi'_{yz}\chi'_{yz} + \chi''_{yz}\chi''_{yz} \right) h_z^2 \right.\\
                    & + \left. 2\left( \chi'_{yy}\chi'_{yz} + \chi''_{yy}\chi''_{yz} \right) h_yh_z\right)
\end{align}
At resonance, substituting Eqs. \eqref{eq:sime_sucep} and \eqref{eq:sucep_yz} into the expression for $<\delta m^2_y(t)> $ yields the simplified form:

\begin{align}
    <\delta m^2_y(t)> & =  \frac{1}{2M^2_s} \left( \left( \chi''_{yy} \right)^2 h_y^2 + \left(\chi'_{yz} \right)^2 h_z^2 \right) \\
                        & = \frac{\chi''_{yy}}{2M^2_s} \left( \chi''_{yy} h_y^2 + \chi''_{zz} h_z^2 \right)\label{eq:dm_y}
\end{align}
Combining this result with Eq. \eqref{eq:m_dmt_2} yields Eq. \eqref{eq:m_dm}, which can be rewritten in terms of the second derivatives of the free energy density:

\begin{align}
     \left< \Vec{m} \times \frac{d\Vec{m}}{dt} \right> &= 2 \omega \sqrt{\frac{\epsilon_{\varphi\varphi}(\pi/2,0)}{\epsilon_{\theta\theta}(\pi/2,0)}} <\delta m^2_y(t)> \vec{x}
\end{align}

In the case of thin magnetic film with perpendicular anisotropy and know in-plane anisotropy it reduce to :
\begin{align}
     \left< \Vec{m} \times \frac{d\Vec{m}}{dt} \right> &= 2 \omega \sqrt{\frac{H_r}{H_r+M_{eff}}} <\delta m^2_y(t)> \vec{x}
\end{align}

\subsection{Energy density for isotropic FM thin film}

In this part, in order to highlight the general results presented in this paper, that only relies for the calibration procedure on two hypothesis :
\begin{itemize}
    \item The cross second derivative evaluated at the equilibrium position is cero.
    \item The damping parameters is sufficiently small in order to only keep the first order term in $\alpha$
\end{itemize}
The first condition is realized in main of the system and imply that the Hessian matrix is diagonal and then $\vec{e_\theta}$ 

Considering a system with out of plane anisotropy and negligible in-plane anisotropy, namely a system with out of plane cylindrical symmetry. It's free density energy is then:
\begin{align}
    \varepsilon(\theta, \varphi)& =-\mu_0 M_s H \left[\cos{\theta_H}\cos{\theta}+\sin{\theta_H}\sin{\theta}\cos{(\varphi-\varphi_H)}\right] \nonumber \\
    & - K_{eff}\cos^2{\theta},
    \label{eqFreeE}
\end{align}
with $K_{eff}=K-\dfrac{\mu_0}{2}M^2_s$, $K$ being the out of plane anisotropy constant in $J/m^3$
Calculating the roots of the first derivatives allow to determine the possible equilibrium postilion and evaluating the second derivatives at the roots :
\begin{align}
     & \epsilon_{\theta\theta}(\theta_{0}, \varphi_{0}) = \mu_{0} M_{S} H_{ext} \sin{(\theta_{0})} + 2K_{eff}\cos{(2\theta_0)} \\ 
    & \epsilon_{\varphi\varphi}(\theta_{0}, \varphi_{0})=\mu_0 M_{S} H_{ext} \sin{(\theta_0)} \\
     & \epsilon_{\theta\varphi}(\theta_{0},\varphi_{0})=0
\end{align}

In this case for an external DC magnetic field applied in the $\vec{x}$ direction $(\theta_H=\pi/2,\varphi_H=0)$, it is straightforward to see that for $K_{eff}<0$ the equilibrium position is equal to $(\theta_0=\pi/2,\varphi_0=0)$. Evaluating the second derivatives at this position it comes out that:
\begin{align}
     & \epsilon_{\theta\theta}(\pi/2, 0) = \mu_{0} M_{S} H - 2K_{eff} \\ 
    & \epsilon_{\varphi\varphi}(\pi/2, 0)=\mu_0 M_{S}H \\
     & \epsilon_{\theta\varphi}(\pi/2, 0)=0
\end{align}
Often one can find the quantity $2K_{eff}$ written in term of $M_{eff}$ with $M_{eff}=-\frac{2K_{eff}}{\mu_0 M_{S}}$. Then it can be rewritten like :
\begin{align}
     & \epsilon_{\theta\theta}(\pi/2, 0) = \mu_{0} M_{S} (H + M_{eff}) \\ 
    & \epsilon_{\varphi\varphi}(\pi/2, 0)=\mu_0 M_{S}H \\
     & \epsilon_{\theta\varphi}(\pi/2, 0)=0
\end{align}

Note : for the equilibrium position along $-\vec{x}$, the equilibrium position is naturally $(\theta_0=\pi/2,\varphi_0=\pi)$, then the values of the secondary derivative are unchanged, but one have to note 
\begin{align*}
    \chi_{yy} & = \chi_{\varphi\varphi}(\pi/2,\pi) \\
    \chi_{zz} & = \chi_{\theta\theta}(\pi/2,\pi) \\
    \chi_{yz} & = \chi_{\varphi\theta}(\pi/2,\pi) \\
\end{align*}
Then the spurious effects have the following parity with respect to the signed of the DC field ($H<0$):
\begin{widetext}
\begin{align*}
   \left<E_y\right> = -\frac{J^x_c}{2M_s} \left( \underbrace{-\rho_{amr} \chi''_{yy}h_y + \rho_{ahe} \chi''_{zz}h_z}_{Symetric} + \underbrace{\rho_{ahe} \chi''_{zy}h_y  + \rho_{amr} \chi''_{yz}h_z}_{AntiSymetric} \right)
\end{align*}
\end{widetext}

To compare to $H>0$, 
\begin{align*}
    \chi_{yy} & = \chi_{\varphi\varphi}(\pi/2,0) \\
    \chi_{zz} & = \chi_{\theta\theta}(\pi/2,0) \\
    \chi_{yz} & = -\chi_{\varphi\theta}(\pi/2,0) \\
\end{align*}

\begin{widetext}
\begin{align*}
   \left<E_y\right> = -\frac{J^x_c}{2M_s} \left( \underbrace{\rho_{amr} \chi''_{yy}h_y + \rho_{ahe} \chi''_{zz}h_z}_{Symetric}+ \underbrace{\rho_{ahe} \chi''_{zy}h_y + \rho_{amr} \chi''_{yz}h_z}_{AntiSymetric} \right)
\end{align*}
\end{widetext}

In conclusion the antisymmetric part is fully odd with field reversal and the symmetric behave even for the AHE part and odd the AMR one.

\subsection{Thermal spurious effects}\label{App:thermal}

Most of the thermal power generated in these geometries arise from the RF absorption in the wave guide, and to a lesser extent the induced RF current inside the heterostructure. In any cases, the joule heating create an heat current density $\vec{J}_Q$ with a DC and a $2\omega$ component. 

In the same way than a charge current density creates an electrical field (a gradient of potential) which will depends on the magnetization direction, an heat current will creates a temperature gradient ($\vec{\nabla}T $) according to the same general equation :

\begin{equation}
   \vec{\nabla}T = k_0 \Vec{J}_Q + k_{div} \left(\Vec{m}.\Vec{J}_Q\right) \Vec{m} + k_{RL} \Vec{m} \times \Vec{J}_Q
\label{Main_Heat_Transport}
\end{equation}
Where, $\Vec{J}_Q$ is the heat current density, $k_0$ is the thermal resistivity, $k_{div}$ the anisotropic magneto-thermal resistance and $k_{RL}$ is the Righi-Leduc coefficient.

In turn, the temperature gradient will translate into a chemical potential gradient, following the magnetic dependent Seebeck coefficients : 

\begin{equation}
   \vec{\nabla} V_{Th} = S_0 \vec{\nabla}T + S_{div} \left(\Vec{m}.\vec{\nabla}T \right) \Vec{m} + S_{ANE} \Vec{m} \times \vec{\nabla}T
\label{Conversion_between_temperature_gradients_and_Chimical_potentiel}
\end{equation}

Where, $S_0$ is the Seebeck coefficient, $S_{div}$ is the Nerst coefficient and $S_{ANE}$ is the Anomalous Nernst coefficient. Then, incorporating \ref{Main_Heat_Transport} into \ref{Conversion_between_temperature_gradients_and_Chimical_potentiel}, it simply gives the same general contributions :

\begin{equation}
   \vec{\nabla} V_{th} = A_0 \Vec{J}_Q + A_{div} \left(\Vec{m}.\Vec{J}_Q \right) \Vec{m} + A_{RL/ANE} \Vec{m} \times \Vec{J}_Q
\label{Main_MagnetoThermic equation}
\end{equation}

Using the same linearization processes than Eq. \ref{eq:m_Jc}, and considering the same equilibrium position along $\vec{x}$ and measuring along $\vec{y}$, the DC contribution of the different thermal gradient will be :

\begin{align}
   \left<E^{th}_y\right> &= A_0 J_Q^y - A_{RL/ANE} J_Q^z + \\
    & A_{div} \left( J_Q^y \left< \delta m_y^2 \right> + J_Q^z \left<\delta m_z \delta m_y \right> \right)
\end{align}
The first term on the right-hand side represents the Seebeck effect and does not take into account magnetic contribution. The second term describes the static contribution of the equilibrium magnetization to the thermal voltage, arising from the combination between the Anomalous Nernst and Righi-Leduc effects. This contribution change of sign when the equilibrium position is reversed resulting in a jump in the measurements at $H_c$. Eventually, the last term correspond to the spurious contribution of the thermal effects at the resonance.

As already calculated (Eq. \ref{eq:dm_y}):
\begin{align*}
    <\delta m^2_y(t)> & =  \frac{1}{2M^2_s} \left( \left( \chi''_{yy} \right)^2 h_y^2 + \left(\chi'_{yz} \right)^2 h_z^2 \right) \\
                        & = \frac{\chi''_{yy}}{2M^2_s} \left( \chi''_{yy} h_y^2 + \chi''_{zz} h_z^2 \right)
\end{align*}
$<\delta m^2_y(t)>$ is symmetric around the resonance, and odd when reversing the static field. In the similar way we compute $\left<\delta m_z \delta m_y \right>$ :
\begin{align}
   \left< \delta m_j(t) \delta m_i(t) \right> =  \frac{1}{2M^2_s} \sum_{k,l} \left( \chi'_{jk}\chi'_{il} + \chi''_{jk}\chi''_{il} \right) h_kh_l
\end{align}
\begin{align}
   \left< \delta m_y(t) \delta m_z(t) \right> & =  \frac{1}{2M^2_s} ( \chi'_{yy}\chi'_{zz} + \chi''_{yy}\chi''_{zz} \\
   & + \chi'_{yz}\chi'_{zy} + \chi''_{yz}\chi''_{zy} ) h_yh_z
\end{align}
Which is also symmetrical with respect to the resonance and even under DC magnetic field. At the resonance condition, and considering the Eq. \ref{eq:sime_sucep} and Eq. \ref{eq:sucep_yz}, we obtain:
\begin{align}
   \left< \delta m_y(t) \delta m_z(t) \right> = 0
\end{align}


\section{Experimental details}

\subsection{Sample and device fabrication}
The heterostructures were grown by room-temperature magnetron sputtering, with the entire stack deposited in situ without breaking vacuum to ensure high interface quality. Devices for SP-FMR and ST-FMR measurements were fabricated via optical lithography. For GAP and SHORT antennas, a four-stage process was used: (1) define the sample bar via Ar-ion etching, (2) deposit Ti(5 nm)/Au(20 nm) contacts, (3) pattern a 200 nm SiO$2$ insulating layer, and (4) deposit Ti(10 nm)/Au(150 nm) to form the microwave antenna and main electrodes. The FULL antenna required only three stages, omitting the contact deposition; these devices were fabricated in parallel and protected with photoresist during the second stage to minimize sample-to-sample variation and ensure identical interfaces across geometries.\\

\end{document}